# Photometry and Spin Rate Distribution of Small-Sized Main Belt Asteroids

D. Polishook [a,b,*] and N. Brosch [a]

[a] *Department of Geophysics and Planetary Sciences, Tel-Aviv University, Israel.*

[b] *The Wise Observatory and the School of Physics and Astronomy, Tel-Aviv University, Israel.*

[*] *Corresponding author: David Polishook, david@wise.tau.ac.il*

**Abstract**

Photometry results of 32 asteroids are reported from only seven observing nights on only seven fields, consisting of 34.11 cumulative hours of observations. The data were obtained with a wide-field CCD (40.5'x27.3') mounted on a small, 46-cm telescope at the Wise Observatory. The fields are located within $\pm 1.5^0$ from the ecliptic plane and include a region within the main asteroid belt.

The observed fields show a projected density of ~23.7 asteroids per square degree to the limit of our observations. 13 of the lightcurves were successfully analyzed to derive the asteroids' spin periods. These range from 2.37 up to 20.2 hours with a median value of 3.7 hours. 11 of these objects have diameters in order of two km and less, a size range that until recently has not been photometrically studied.

The results obtained during this short observing run emphasize the efficiency of wide-field CCD photometry of asteroids, which is necessary to improve spin statistics and understand spin evolution processes. We added our derived spin periods to data from the literature and compared the spin rate distributions of small main belt asteroids (5>D>0.15 km) with that of bigger asteroids and of similar-sized NEAs. We found that the small MBAs do not show the clear Maxwellian-shaped distribution as large asteroids do; rather they have a spin rate distribution similar to that of NEAs. This implies that non-Maxwellian spin rate distribution is controlled by the asteroids' sizes rather than their locations.

**Motivation and background**

Photometry of asteroids has been performed for more than a century, since von Oppolzer suggested that the observed light variability of *(433) Eros* is due to rotation of an object of irregular shape (Gehrels 1979). Today, photometry of asteroids is a primary tool to investigate their physical properties such as rotation period, shape, obliquity, size and structure. By now (September 2008) more than 3,400 asteroids have published lightcurves (see the Web site of A.W. Harris and B. D. Warner http://www.minorplanetobserver.com/astlc/LightcurveParameters.htm), most of them belong to the main belt of asteroids (MBAs), about 440 are near-Earth asteroids (NEAs) and a few dozen are Trojans and trans-Neptunian objects (TNOs). This large mass of information is beneficial and critical to study spin distributions of asteroids and to understand asteroids' evolution processes.

Over the years, the spin rate distribution of MBAs was checked by different authors for an increasing number of asteroids with spin measurements, while the diameters of the targeted asteroids decreased as the sensitivity of the instruments increased. Binzel et al. (1989) summarized the spin rate values of 375 asteroids and found that the spin rate distribution of asteroids as large as 125 km and larger fits a Maxwellian distribution, while smaller asteroids with 125>D>50 km fit a superposition of two Maxwellian distributions. Smaller asteroids deviate from a Maxwellian or from any linear combinations of a small number of Maxwellian distributions. Fulchignoni et al. (1995) analyzed the rotation rates of 516 asteroids and found that while the spin rate distribution of "large" asteroids with D>50 km fits a Maxwellian distribution, that of the "small" asteroids (D<50 km) fits a linear combination of Maxwellian distributions of three groups of asteroids: fast rotating (mean period value of 3.9 hours), mid-rotators (8.6 hours; which matches the distribution of the "large" asteroids) and slow-rotators (23.9 hours). Fulchingoni et al. (1995) also searched for deviations within the different taxonomic groups and found small differences in the sense that denser asteroids (M-type) rotate faster, on average, than less dense asteroids (C-type and P-type), while the S-type and D-type are in-between.

The study of Pravec & Harris (2000) included 750 asteroids, 94 of which are NEAs. The short geocentric distance of NEAs enabled Pravec & Harris to stretch farther the boundary between "small" and "large" asteroids: they divided the dataset into large asteroids with diameters D>40 km, intermediate (40>D>10 km) and small (10>D>0.15 km). Pravec and Harris found that the spin rate distribution of the large asteroids fits a Maxwellian; the small asteroids show a completely non-Maxwellian behavior; while the intermediate group is a transitional region where the large and small asteroid groups overlap. The existence of many fast and slow rotators among the small asteroids suggests that other mechanisms rather than collisions have shaped their spin rate distribution. The fact that many of the small objects with resolved rotation periods are NEAs, that orbit the Sun in regions with different conditions than those of MBAs, brings up the question whether small-sized MBAs have the same spin rate distribution as their NEAs counterparts (Pravec et al. 2002, Binzel et al. 2002). While disruptive and non-disruptive collisions are supposed to be the primary mechanisms that shape the spin rate distribution of MBAs, NEAs will additionally be affected by tidal forces following close encounters with the terrestrial planets (Scheeres et al., 2004). However, NEAs are former MBAs that were injected into the inner Solar System and survive in this state only for $\sim 10^7$ years (Bottke et al. 2002). Therefore, the spin distribution of the NEAs should be similar to the spin distribution of the MBAs, excluding the additional contributions by the tidal forces of the terrestrial planets and by an enhanced YORP effect which inversely depends on the square of the heliocentric distance. Are these contributions seen in the spin rate distribution derived by Pravec & Harris (2000) for small asteroids (10>D>0.15 km) that shows non-Maxwellian behavior, or is this the result of the small size of the asteroids? Could these contributions be seen only in NEAs' spin rate distributions but not in that of small MBAs? Do small MBAs have a Maxwellian-shaped spin rate distribution?

Here we present observations related to the spin periods of small MBAs obtained at the Wise Observatory, and combine those with published values for similar objects to compare spin rate distribution of small MBAs to larger groups of MBAs and to similar-sized NEAs.

**Observations, reduction, measurements and calibration**

Observations were performed with a new reflector telescope at the Wise Observatory, an 18" Centurion telescope (referred to as *C18;* see Brosch et al. 2008 for a description of the telescope and performance). An SBIG ST-10XME CCD was used at the f/2.8 prime focus. This CCD covers a field of view of 40.5'x27.3' with 2184x1472 pixels, with each pixel subtending 1.1 arcsec, and is used in white light with no filters. The asteroids were observed while crossing a single field per night, thus the same comparison stars were used while calibrating the images (see below).

The observations took place on October 4, 5, 6 and 31, and on November 2, 3 and 5, all in 2007. On six of the seven nights the telescope was aimed at the NEA *(2212) Hephaistos* for an average of 5.5 hours per night, while it was crossing the main belt at a heliocentric distance of 2.74 to 2.88 AU at a range of ±1.5 degrees from the ecliptic plane. As a result, 7.2 MBAs on average were included in each field and were photometrically measured, in addition to the primary target *(2212) Hephaistos*. Details of the observed fields are summarized in Table I. An example of the dense fields is shown in Fig. 1 and demonstrates the field size while exhibiting our record of 11 asteroids imaged at once. On the last night (November 5, 2007) we performed a follow-up observation on one of the asteroids, *(36405) 2000 OB$_{48}$*, which was observed during a previous night. All together, we observed 32 different asteroids excluding *(2212) Hephaistos.*

*[put **Table I** about here]*

*[put **Fig 1** about here]*

To achieve a point-like FWHM within the observation conditions (averaged angular velocity of ~0".01 per sec, FWHM of ~2.5 pixels due to seeing) an exposure time of 150 seconds was chosen (120 seconds on November 5). The observational circumstances are summarized in Table II, which lists the asteroid's designation, the observation date, the time span of the observation during that night, the number of obtained images, the object's heliocentric distance (r), geocentric distance (Δ), phase angle (α), and the Phase Angle Bisector (PAB) ecliptic coordinates (L$_{PAB}$, B$_{PAB}$ - see Harris et al. (1984) for the definition and ways of calculating these parameters). In addition, the object's mean observed magnitude following calibration with standard stars (see calibration method below) is listed for each night.

*[put **Table II** about here]*

The images were reduced in a standard way using bias and dark subtraction, and division by a normalized flatfield image. Times were corrected to mid-exposure. We used the IRAF *phot* function for the photometric measurements. Apertures with a radius of four pixels were chosen, in order to minimize photometric errors. The mean sky value was measured using an annulus with an inner radius of 10 pixels and 10 pixels wide around the asteroid.

After measuring, the photometric values were calibrated to a differential magnitude level using ~350 local comparison stars that were measured on every image by the same method as the asteroid. A photometric shift was calculated for each image compared to a good reference image, using the local comparison stars. Variable stars were removed at a second calibration iteration leaving an average of 160±27 local comparison stars per image. The brightness of these stars remained constant to ±0.02 mag..

The instrumental photometric values were calibrated to standard magnitudes using 44 measurements of Landolt equatorial standards (Landolt 1992). These were observed at air masses between 1.1 to 2.5, while switching between the asteroid fields that included the same local comparison stars used for the relative calibration. This calibration observation was done on November 8, 2007 under photometric conditions. The extinction coefficients for the photometric nights, including the zero point, were obtained using the Landolt standards after measuring them in an aperture with a ten-pixel radius. From these, the standard magnitudes of the local comparison stars of each field were derived, followed by calculating the magnitude shift between the daily weighted-mean magnitude and the catalog magnitude of the comparison stars. This magnitude shift was added to the data of the relevant field and asteroids, and introduced an additional error of ~0.02-0.03 mag. due to the observational systematic errors of the standards stars and the comparison stars, and due to uncertainties in matching the photometric coefficients. Since the images were obtained in white light, which corresponds to a wide $R$ band, we used the Cousins $R$ magnitudes of the Landolt standards for the calibration. We note that the accepted range of asteroidal colors was included in the wide color range of the observed standards (-0.03 ≤ *B-V* ≤ 1.1), and the color term of our transformation from instrumental to *Cousins R* was smaller than 0.01 mag..

The asteroid magnitudes were corrected for light travel time and were reduced to a 1 AU distance from the Sun and the Earth to yield reduced R values (Bowell et al. 1989). Astrometric solutions were obtained using *PinPoint* ([www.dc3.com](www.dc3.com)) and some positions were reported to the MPC and checked with its web database. Six objects of the 32 observed asteroids were not registered on the MPC web site. For one of them we managed to obtain a follow-up measurement on October 15 and received a new designation from the minor planet center – *2007 TD$_{106}$*.

**Analysis**

To determine the lightcurve period and amplitude, the data analysis included folding all the calibrated magnitudes to one rotation period at zero phase angle using two basic techniques: the Fourier series for determining the variability period(s) (Harris and Lupishko 1989) and the *H-G* system for calibrating the phase angle influence on the magnitude (see appendix in Bowell et al. 1989). Together, these two relations constrain the rotation period (P), the Fourier coefficients ($B_n$, $C_n$), the absolute magnitude in the R filter ($H_R$) and the slope parameter (G):

$$R(1,\alpha,t) + 2.5\cdot\log[(1-G)\cdot\Phi_1(\alpha)+G\cdot\Phi_2(\alpha)] = H_R + \sum_{n=1}^{m} \{B_n\cdot\sin[(\frac{2n\pi}{P})\cdot(t-t_0)]+C_n\cdot\cos[(\frac{2n\pi}{P})\cdot(t-t_0)]\} \quad (1)$$

Here R(1,α,t) is the reduced magnitude at any phase angle (α) and time (t). To simplify Eq. 1, a guess for the frequency f ($=\frac{2n\pi}{P}$) and the slope parameter G is adopted. This yields a linear equation that is easily solved using least squares. Most of the objects were analyzed using only two harmonics to fit the data to a simple model.

In these cases the model may not fit some features of the lightcurve. In a few cases [*(36405) 2000 OB$_{48}$, (132114) 2002 CE$_{224}$, 2002 CP$_{33}$, 2003 QU$_{50}$*), where many measurements with small errors were available, we fitted the data using a model of 6 harmonics. Since most asteroids described here were not observed over a wide phase angle range, we used a default value of G=0.15±0.12. The error of the G slope in this general case is the standard deviation of many G values found in the study of Lagerkvist and Magnusson, 1990 (see their Table II, third column). For asteroids observed at small phase angles, this G slope error introduce an error of about 0.08 magnitudes to the absolute magnitude H$_R$. For two asteroids the slope parameter G was derived and their phase curves are shown using the mid-amplitude values of the lightcurves.

Constraints on asteroid shapes were derived from the amplitude of the lightcurve, assuming a triaxial body with a ≥ b ≥ c and object rotation about the c axis (Harris & Lupishko 1989). At the first step, the amplitude A(α) was calibrated to zero phase angle A(0$^0$) using the method of Zappala et al. (1990):

$$A(0^0) = A(\alpha) / (1 + m \cdot \alpha) \qquad (2)$$

where α is the phase angle and m is the slope that correlates the amplitude to the phase angle. We used the average value of m=0.023 deg$^{-1}$ found by Zappala et al. (1990) from 27 different measurements. Using the calibrated amplitude A(0$^0$), we calculated a minimum value for the axial ratio a/b using:

$$A(0^0) = 2.5 \cdot \log\left(\frac{a}{b}\right)_{min} \qquad (3).$$

When the lightcurve is not complete we provide minimal limits for the period and amplitude, assuming a two-peak lightcurve, and give a rough estimate for the absolute magnitude H$_R$. An extended description of the analysis method appears in Polishook & Brosch (2008).

**Results**

We did not find reports of previous photometric measurements for the asteroids described here [except for the test case of *(698) Ernestina*]. We separate this section into two sub-sections: one deals with those objects for which we have data of sufficient quality to derive full lightcurves, in which we analyze and present results for individual asteroids. The other sub-section describes the results for those objects where the data is either not long enough, or has a low signal-to-noise ratio. Since it is not possible to yield a full characterization of the periodic lightcurves for these objects, we provide only limited information for each individual asteroid.

*Asteroid lightcurves*

Thirteen of the 32 asteroids were observed sufficiently to allow the derivation of conclusive periods. These are listed in Table III, which includes the asteroid name, rotation period, reliability code (according Harris et al. 1999), photometric amplitude, absolute magnitude $H_R$ and slope parameter G. We exclude *(2212) Hephaistos* from the list and will publish its lightcurve in a subsequent paper. Our photometry results for the asteroid *(698) Ernestina,* measured here as a test case, are compared with the results of Ivanova et al. (2002).

[put **Table III** about here]

*(698) Ernestina*

To verify our data collection and reduction methods, we observed a well-known asteroid with a published rotation period – *(698) Ernestina*. This is a 10.7 mag. asteroid (at the IAU H magnitude scale) that orbits the Sun at the inner edge of the outer main belt (a=2.870 AU). The observations were done at different times (November 6 & 7, 2005 and January 21 & 22, 2006) and all the stages of measurement, calibration and analysis were identical to those done for the other observed asteroids. Table II shows the observational circumstances of *(698) Ernestina*. Ivanova et al. (2002) observed *(698) Ernestina* on 13 February 2002 and obtained a synodic rotation period of 5.07±0.01 hours with an amplitude of 0.637±0.027 mag.. Unpublished observations done on Jan. 14 and 20 , 2002 and on Mar. 9, 2002 by L. Bernasconi (reported at http://obswww.unigh.ch/~behrend/page_cou.html) showed a rotation period of 5.03653±0.00004 hours with an amplitude of 0.690±0.015 mag..

Using Fourier analysis, we derived a rotation period of 5.0431±0.0002 hours, bracketed by the two published values, and two amplitudes: 0.35±0.03 and 0.26±0.03 mag. indicating that *(698) Ernestina* has an asymmetric shape (see lightcurve in Fig. 2). The observations of November 2005 showed that one of the peaks was fainter by 0.06 mag. than the observations of January 2006, probably representing the changed phase angle of the asteroid. Even though the accumulated data on the changing amplitude constrain somewhat the spin axis orientation, we could not derive a preferred fit to the pole orientation and more observations at different apparitions are needed to derive it.

 *[put **Fig 2** about here]*

Given the wide range of observed phase angles, from $3.3^0$ to $20.6^0$, we found a good fit for the different phase slope parameters (Fig. 3): an absolute magnitude ($H_R$) of 10.75±0.02 mag., and a slope parameter (G) of 0.21±0.02 on the *H-G* system.

*[put **Fig 3** about here]*

*(16345) 2391T-3*

Located within the outer MB (a=2.918 AU), *(16345) 2391T-3* was observed on November 3, 2007. We derived a rotation period of 2.9±0.2 hours and amplitude of 0.25±0.05 mag. using Fourier analysis (Fig. 4). Even though the lightcurve was fully covered, we mark the reliability code as 2+, due to the low amplitude compared to the S/N. The absolute magnitude $H_R$ is 14.89±0.07 mag., assuming G=0.15.

*[put **Fig 4** about here]*

*(36405) 2000 OB$_{48}$*

This inner MB object (a=2.285 AU) was observed on October 4 2007 and showed an almost complete lightcurve with very high amplitude and low photometric error; we reobserved it on November 5 to obtain a better sampled lightcurve. We derived a rotation period of 6.6±0.3 hours using Fourier analysis and obtained two amplitudes: 0.82±0.02 mag. and 0.62±0.02 mag. (Fig. 5). After calibrating the amplitude to zero phase angle, these represent a minimal a/b ratio of 1.75±0.03 and 1.53±0.03 and may suggest that *(36405) 2000 OB$_{48}$* has an asymmetric shape. Note that the deepest minimum was not observed completely and it might even be deeper and the asteroid's shape can be more elongated. A fit to the *H-G* system yields an absolute magnitude of $H_R$=15.27±0.04 mag. and G=0.07±0.04 (Fig. 6).

*[put **Fig 5-6** about here]*

*(42535) 1995 VN$_9$*

At the far edge of the inner main belt (a=2.515 AU), this asteroid was observed on October 31, November 2 and 3, 2007. Even though the data are noisy, a rotation period of 2.37±0.01 hours and amplitude of 0.3±0.1 mag. were derived using

Fourier analysis (Fig. 7). Even though the lightcurve was covered more than twice, we mark the reliability code as 2+, due to the low amplitude compared to the S/N. The absolute magnitude $H_R$ is 16.0±0.1 mag. assuming G=0.15.

*[put **Fig 7** about here]*

*(75598) 2000 AY$_{23}$*

This asteroid, located within the central MB (a=2.633 AU), was observed only on November 2, 2007, but showed more than three maxima and three minima in its lightcurve although ~25% of the data points were lost due to a short projected angular distance to a very bright star. Using Fourier analysis we derived a rotation period of 3.05±0.05 hours and amplitude of 0.33±0.03 mag. (lightcurve at Fig. 8). Assuming G=0.15, the estimated absolute magnitude $H_R$ of *(75598) 2000 AY$_{23}$* is $H_R$=15.41±0.07 mag..

*[put **Fig 8** about here]*

*(106836) 2000 YG$_8$*

This asteroid, located within the inner MB (a=2.381 AU), was observed on October 31 and November 2, 2007. A long period of 20.1±0.2 hours was derived using Fourier analysis under the assumption of a two-peak lightcurve (Fig. 9). The amplitude is 0.65±0.03 mag., which after calibration correlates with a minimal a/b ratio of 1.72±0.05. Note the possible presence of a systematic dimming at phase ~0.8; this requires confirmation by subsequent observations. The absolute magnitude $H_R$ is 16.53±0.08 mag. assuming G=0.15.

*[put **Fig 9** about here]*

*(129997) 1999 VH$_{28}$*

Located within the far end of the inner main belt (a=2.548 AU), *(129997) 1999 VH$_{28}$* was observed on November 2 and 3, 2007. Using Fourier analysis a rotation period of 8.8±0.2 hours was derived with an amplitude of 0.25±0.03 mag. (Fig. 10). Assuming G=0.15, the absolute magnitude H$_R$ is 15.74±0.07 mag.. Note that the noisy data points fit also a solution of 6.59±0.09 hours (with the same chi-square score), but this model fits an irregular lightcurve with high ratio between the two amplitudes. Therefore we chose the simpler model of P=8.8±0.2 hours.

*[put **Fig 10** about here]*

*(132114) 2002 CE$_{224}$*

This inner MB asteroid (a=2.352 AU) was observed on three consecutive nights on October 4, 5 and 6, 2007. A rotation period of 5.672±0.003 hours is derived using Fourier analysis (lightcurve in Fig. 11). The high amplitude of 1.25±0.05 mag. correlates with a high a/b axis ratio of 2.4±0.1, after calibrating the amplitude to zero phase angle [*(132114) 2002 CE$_{224}$* was observed at a phase angle of ~12.5$^0$]. The absolute magnitude H$_R$ is 16.7±0.2 mag..

*[put **Fig 11** about here]*

*(168714) 2000 JQ$_{10}$*

This asteroid, located in the inner MB (a=2.210 AU) was observed on November 3, 2007. A rotation period of 2.5±0.2 hours was derived using Fourier analysis, with an amplitude of 0.15±0.05 mag. (Fig. 12). Even though the lightcurve was fully covered, we mark the reliability code as 2+, due to the low amplitude compared to the S/N. The absolute magnitude $H_R$ is 17.01±0.09 mag. assuming G=0.15.

*[put **Fig 12** about here]*

*(168847) 2000 $UU_{34}$*

This inner MB asteroid (a=2.311 AU) was observed on November 2, 2007. Although this was a low S/N observation (reliability code of 2+), a rotation period of 2.9±0.1 hours was derived using Fourier analysis with an amplitude of 0.30±0.05 mag. (Fig. 13). The absolute magnitude $H_R$ is 16.79±0.08 mag. assuming G=0.15.

*[put **Fig 13** about here]*

*(168904) 2000 $WY_{189}$*

This asteroid, located within the inner MB (a=2.321 AU), was observed only on November 2, 2007, but showed more than four peaks and four valleys in its lightcurve (Fig. 14). We derived a rotation period of 3.335±0.009 hours and an amplitude of 0.36±0.05 mag.. Assuming G=0.15, the estimated absolute magnitude $H_R$ of *(168904) 2000 $WY_{189}$* is $H_R$=17.58±0.09 mag..

*[put **Fig 14** about here]*

*2002 $CP_{33}$*

This asteroid, located within the inner MB (a=2.200 AU) was observed on November 2 and 3, 2007. A period of 4.78±0.02 hours was derived using Fourier analysis (lightcurve shown in Fig. 15). A high amplitude of 0.8±0.1 mag. was measured, which correlates, after calibration, with a minimal a/b ratio of 1.9±0.2. Note that the first peak (at the folded lightcurve) is not as symmetric as the second peak and seems to exhibit a small drop in the brightness at phase ~0.2. This drop in brightness, which is clearly seen on three out of five peaks at the observed lightcurve (lower panel), might represent an irregular or non-convex shape. The absolute magnitude $H_R$ is $H_R=17.2\pm0.1$ mag., assuming a G=0.15.

*[put **Fig 15** about here]*

*2003 QU$_{50}$*

This inner MB asteroid (a=2.390 AU) was observed on two consecutive nights on October 4 and 5, 2007. We derived a rotation period of 4.14±0.01 hours and amplitude of 0.6±0.1 mag. using Fourier analysis (lightcurve shown in Fig. 16). The absolute magnitude is $H_R=16.3\pm0.2$ mag. assuming G=0.15.

*[put **Fig 16** about here]*

*2007 TF$_5$*

This recently discovered asteroid located in the central MB (a=2.607 AU), was observed on October 31, 2007. A short rotation period of 2.5±0.2 hours and amplitude of 0.4±0.1 mag. were derived by Fourier analysis (Fig. 17). Reliability code was marked as 2+, due to the low amplitude compared to the S/N. Assuming G=0.15, an absolute magnitude $H_R$ of 16.1±0.1 mag. was calculated.

*[put **Fig 17** about here]*

*Partial results*

The following asteroids were only observed for a relatively short period and their lightcurves are only partially covered, thus their rotation periods could not be fully resolved. The relevant results are summarized in Table IV. We show some lightcurves with second order polynomial fits that yield minimal spin periods and minimal photometric variability. In most cases, when the asteroid was observed during only one night, the minimal spin period is twice the observed time due to the double nature of asteroids lightcurves. We also estimated a rough absolute magnitude $H_R$ by projecting the mean reduced R magnitudes to zero phase angle assuming G=0.15. Since the full amplitude was not covered in these cases, an additional error of 0.4 mag. is introduced. We remark that more observations are needed to exactly resolve the spin periods of these asteroids.

*[put **Table IV** about here]*

*(8549) Alcide*

Located at the inner MB (a=2.438 AU), the asteroid *(8549) Alcide* was observed on November 5, 2007. Even though its lightcurve was not fully covered and is showing less than half a cycle (Fig. 18), we can determine that its spin period is at least 2.8 hours with a minimum amplitude for variability of 0.25±0.03 mag.. Using the mean value of the reduced data points and assuming a G=0.15, an $H_R$ value of 14.3±0.4 mag. can be estimated.

*[put **Fig 18** about here]*

*(70517) 1999 $TU_{105}$*

This asteroid, located within the far side of the inner MB (a=2.527 AU), was observed on October 31, 2007 for 5.62 hours. Through this time range *(70517) 1999 $TU_{105}$* showed a minimum in its lightcurve (Fig. 19), suggesting a minimal period of 11.2 hours and a minimum variability of 0.21±0.03 mag.. An $H_R$ value of 15.7±0.4 mag. can be derived assuming G=0.15.

*[put **Fig 19** about here]*

*(106864) 2000 YL$_{27}$*

Even though not fully covered, the lightcurve of *(106864) 2000 YL$_{27}$* fits most nicely a rotation period of 4.9±0.1 hours. However we cannot give a precise value until this inner MB asteroid (a=2.390 AU) will be observed for at least one full rotation. From the data points obtained on October 5 and 6 (Fig. 20) we can suggest a lower limit to the spin period of 4.5 hours. The minimal variability amplitude is about 0.3 mag.. Assuming G=0.15, H$_R$ is 16.2±0.4 mag..

*[put **Fig 20** about here]*

*(166243) 2002 GL$_{13}$*

The observations of this inner main belt asteroid (a=2.354 AU) took place on November 5. The lightcurve of *(166243) 2002 GL$_{13}$* (Fig. 21) suggests that only a short segment of its rotation period was observed. A minimal variability period of 2.8 hours and a minimum variability amplitude of 0.2 mag. is suggested. Using the mean value of the reduced data points and assuming G=0.15, an H$_R$ value of 16.1±0.5 mag. is estimated.

*[put **Fig 21** about here]*

*2005 EJ$_{141}$*

The rotation period of this inner MB asteroid (a=2.360 AU) seems to be long after showing, during almost 6 hours of observations (on October 4, 2007) only a short part of an ordinary lightcurve. We present in Fig. 22 the data from which a minimum spin period of 21 hours, a minimum variability amplitude of 0.4±0.1 mag.

and an absolute magnitude $H_R$ of 17.9±0.5 mag. can be derived, but note that the real values can be much different than these limits.

*[put **Fig 22** about here]*

*Inconclusive results*

Some asteroids were too faint, or their data were too noisy, to derive a solid spin period result or even limiting values for their spins. Table V describes the number of data points collected and the derived absolute magnitudes $H_R$ using the mean value of the observations and assuming G=0.15.

*[put **Table V** about here]*

**Discussion and conclusions**

The observed asteroids are presumably some of the smallest MBAs for which rotational properties have been investigated. They enlarge by ~30% the Harris & Warner's database of rotational properties of small main belt asteroids (for D<2 km). As a result, we can combine our sample with published spin period data to produce a spin rate distribution for the small MBAs, and can compare it to the spin rate distribution of larger MBAs. Moreover, we can compare the spin rate distributions of small MBAs to that of NEAs that consist of small objects only, to attempt the detection of any spin differences that could have developed due to different environmental conditions.

To estimate the diameters of asteroids we follow the simple rule of transforming the absolute magnitude H to diameter D (in km), assuming an albedo *Pv* (Pravec et al. 1998):

$$D = \frac{1329}{\sqrt{Pv}} 10^{-0.2H} \qquad (4)$$

While Eq. 4 is used for magnitudes in the V-band, the observations with the *C18* are taken with no filter. This, as mentioned above and as detailed in Brosch et al. (2008) corresponds to a wide-R band. Therefore, we assumed a color of V-R=0.45 mag. to translate our measured $H_R$ to $H_V$ magnitudes ($H_V$ = H), in consistent way to the method of Pravec et al. (1998). A color variation of ±0.1 mag. may change the diameter estimate by less than five percent, thus the observed asteroids would still be in the size range of ~1-2 km and such color variations can be neglected.

Albedo values run from ~0.05 for dark carbonaceous asteroids up to ~0.5 for the shiniest "metal" surfaces of the E-type asteroids. To use a more reliable albedo value we assumed that dark C-type asteroids are more common in the outer regions of the MB than S-types. Although recent studies differ on the heliocentric distance at which the transition takes place (Bus & Binzel 2002, Mothé-Diniz et al. 2003), we followed the estimate that S-type asteroids are the majority at the inner main belt (Gradie & Tedesco 1982). For simplicity, we chose an albedo of 0.18 for asteroids from the inner main belt (a<2.6 AU), 0.10 for asteroids from the central main belt (2.6<a<2.7 AU) and 0.058 for asteroids from the outer main belt (a>2.7 AU). For consistency, we follow the Harris and Warner region definition as appears in their lightcurve database. Asteroids from a distinct family or group (like the Flora group) received the average albedo of the group. Table VI presents the assumed albedo and calculated diameter for each asteroid.

We did not use a higher albedo for the inner MBAs in order to retain a conservative, not-too-small, estimate for the diameters of these asteroids. One should, therefore, keep in mind that these asteroids might even be smaller than calculated here. In any case, Pravec & Harris (2000) estimated the error of the size calculation using this method as a factor of 1.5 to 2.

*[put **Table VI** about here]*

We used spin periods from the updated list (March 2008) of Harris & Warner (http://www.minorplanetobserver.com/astlc/LightcurveParameters.htm) and added the results derived here, for a total of 1445 spin period values of MBAs, not including those of the *Hirayama* families such as *Flora*, *Phocaea* and *Eos*. Asteroids with inner orbits compared to the main belt, such as the Mars-orbit crossers and the *Hungaria* group, were also excluded as done for NEAs, *Trojans*, *Centaurs* and *TNOs*. We also excluded spin values with quality code of 1, meaning that these are based only on fragmentary lightcurves and could be completely wrong. We followed Pravec and Harris (2000) by dividing the asteroids into groups by size: D>40 km (L = large), 40>D>10 km (M = medium) and 10>D>0.15 km. Given our larger data set, we divided the last group further into 10>D>5 km (S = small) and 5>D>0.15 km (VS = very small) so the asteroids in the last group are equivalent in size to NEAs. Figures 23-26 present the comparison between the spin rate distribution of the small MBAs group to those of larger MBAs and the spin rate distribution of the NEAs.

*[put **Fig 23-24-25-26** about here]*

The histograms (Fig. 23) show that the VS group (5>D>0.15 km) does not exhibit the clear Maxwellian-shaped distribution as the L group (D>40 km). The M group (40>D>10 km, Fig. 24) is a transitional region where the large and small asteroid groups overlap. A Kolmogorov-Smirnov (KS) test with a five percent significance level rejected the null hypothesis that the VS distribution is drawn from the same underlying continuous population as the L or the M distribution (P-value<0.0001). The small MBA distribution is observationally biased against slow rotators (due to lack of sufficient observations to cover a significant fraction of a rotation period), which explains the lack of measurements in this regime. This distribution is also biased against fast rotators (because the sampling is too sparse compared to the brightness variability period), but nevertheless shows more objects in the fast spinning area, supporting the conclusion of a non-Maxwellian distribution of small asteroids. This is consistent with the study of Pravec & Harris (2000) that compared spin rate distribution of all small asteroids (10>D>0.15 km) to that of bigger asteroids. On top of these, we can add two conclusions:

*i)* The non-Maxwellian distribution of small asteroids is common between small MBAs and small NEAs (Fig. 26), thus this non-Maxwellian spin rate distribution is controlled by the asteroids' sizes rather than their locations. This conclusion might limit the effect of a short heliocentric distance as an important factor of the YORP effect, or at least set a minimum limit to its timescale of $10^7$ years, the lifetime of asteroids as NEAs. However, more photometric observations of small MBAs are needed to detect subtle differences between the two histograms. Such differences, if they exist, can help us study the specific effects of the tidal forces of the terrestrial planets and the YORP effect.

*ii)* While the spin rate distributions of the S and VS MBAs (10>D>5 km and 5>D>0.15 km) match almost perfectly (Fig. 25), there are some fast rotators (10-13 revolutions per day) in the VS group (5>D>0.15 km) that hardly show up in the histogram of the S group. This difference might suggest that when a body is smaller than five kilometers another physical mechanism (such as the YORP effect) starts to efficiently affect the asteroid's spin rate. Even though this interesting conclusion should be examined further with more spin periods of small MBAs, both populations are subject to the same observational biases against the measurements of fast rotators, thus we do not anticipate finding more fast rotators in the S group compared to the VS group.

These conclusions should be confronted with the questions marks put by observational biases especially when we bear in mind that only 13 out of 32 asteroids were analyzed successfully while minimal limits were found for 5 asteroids and for 14 other asteroids the photometric variability was hidden by the low S/N. It is also possible that these 14 asteroids are slow rotators and their brightness did not vary dramatically when they were observed. To correct the observational biases we consider four extreme scenarios: i) The 14 asteroids have the same spin distribution as the VS MBA group. ii) The 14 asteroids have the same spin distribution as the L MBA group. iii) All 14 asteroids are fast rotators. iv) All 14 asteroids are slow rotators. Scenarios i and iii do not affect, and possibly enhance our conclusions. To debias our database of small MBAs in order to examine scenarios ii and iv, we should remember that the 14 asteroids with inconclusive data, should be compared with the solved 13 asteroids of this survey alone and not with literature data of small MBAs. Moreover, as mentioned above, this comparison is done only for main belt asteroids and not for objects from different Hirayama families. This reduces the two lists to nine asteroids with solved periods and nine asteroids with unsolved periods. Therefore, while examining scenario ii, the debiased spin rate distribution of the VS group (5>D>0.15 km) includes one half of the data distributed as the L group, and the other half of the data distributed as the VS group. The comparison between this debiased distribution to the spin rate distribution of large MBAs (Fig. 27) clearly shows that the distribution of fast rotators among the small MBA population is significant with respect to the rest of the bins, and the debiased distribution is not a pure Maxwellian, supporting the conclusions of this paper. A Kolmogorov-Smirnov (KS) test with a five percent significance level rejected the null hypothesis that the debiased VS distribution is drawn from the same underlying continuous population as the L distribution (P-value<0.003). Scenario iv is even more radical: distributing half of the small MBAs in the spin rate regime of less then 2 revolutions per day, displays a non-Maxwellian distribution (P-value<0.003 at the KS test). These two scenarios show that even if extreme situations are taken into consideration, the spin rate distribution of small MBAs is non-Maxwellian and is different than those of large MBAs.

*[put **Fig 27** about here]*

To further support the conclusions of this study we continue to photometrically measure the main belt and we encourage the community to obtain more spin values of small MBAs until a statistically significant result would be reached.

**Acknowledgements**

DP is grateful for an *Ilan Ramon* doctoral scholarship from the Israel Space Agency (ISA) and the Ministry of Science, Sports and Culture. We thank the Wise Observatory staff for their continuous support. The research was supported by the Israeli Ministry of Science, Culture and Sport that created a National Knowledge Center on NEOs and asteroids at the Tel-Aviv University.

**References**

Binzel, R. P., Farinella, P., Zappalà, V. and Cellino, A., 1989. Asteroid Rotation Rate – Distribution and Statistics. In *Asteroids II* (R. P. Binzel, T. Gehrels and M. S. Matthews, Eds.), pp. 524-556. Univ. of Arizona Press, Tucson.

Binzel, R. P., Lupishko, D. F., Di Martino, M., Whiteley, R. J. and Hahn, G. J., 2002. Physical Properties of Near-Earth Objects. In *Asteroids III* (W.F. Bottke Jr. et al. eds.), pp. 255-271, Univ. of Arizona, Tucson.

Bottke, W. F., Morbidelli, A., Jedicke, R., Petit, J., Levison, H., Michel, P. and Metcalfe, T., 2002. Debiased orbital and absolute magnitude distribution of the Near-Earth Objects. *Icarus* **156**, 399-433.

Bowell, E., Hapke, B., Domingue, D., Lumme, K., Peltoniemi, J. and Harris, A. W., 1989. Application of photometric models to asteroids. In *Asteroids II* (R. P. Binzel, T. Gehrels and M. S. Matthews, Eds.), pp. 524-556. Univ. of Arizona Press, Tucson.

Brosch, N., Polishook, D., Shporer, A., Kaspi, S., Berwald, A. and Manulis, I., 2008. The Centurion 18 telescope of the Wise Observatory. *Astrophys. Space Sci.* in press.

Bus, S. J. and Binzel, R. P., 2002. Phase II of the small Main Belt asteroid spectroscopic survey. A feature based taxonomy. *Icarus* **158**, 146-177.

Fulchignoni, M., Barucci, M. A., Di Martino, M. and Dotto, E., 1995. On the evolution of the asteroid spin. *Astron. Astrophys.* **299,** 929–932.

Gehrels 1979. The Asteroids: History, Surveys, Techniques, and Future Work. In *Asteroids* (T. Gehrels, ed.), pp. 3-24, Univ. of Arizona, Tucson.

Gradie J. and Tedesco, G., 1982. Compositional Structure of the Asteroid Belt. *Science* **216**, 1405-1407.

Harris, A. W., Young, J. W., Scaltriti, F. and Zappalà, V., 1984. Lightcurves and phase relations of the asteroids 82 Alkmene and 444 Gyptis. *Icarus* **57**, 251-258.

Harris, A. W. and Lupishko, D. F., 1989. Photometric lightcurve observations and reduction techniques. In *Asteroids II* (R. P. Binzel, T. Gehrels and M. S. Matthews, Eds.), pp. 39-53. Univ. of Arizona Press, Tucson.

Harris, A. W., Young, J. W., Bowell, E. and Tholen, D. J., 1999. Asteroid lightcurve observations from 1981 to 1983. *Icarus* **142**, 173-201.

Ivanova, V. G., Apostolovska, G., Borisov, G. B. and Bilkina, B. I., 2002. Results from Photometric Studies of Asteroids at Rozhen National Observatory, Bulgaria. In proceedings of *Asteroids, Comets, Meteors (ACM 2002)*, 29 July – 2 August, 2002. Technical University Berlin. 505-508.

Lagerkvist, C.-I. & Magnusson, P., 1990. Analysis of asteroid lightcurves. II - Phase curves in a generalized HG-system. *Astron. Astrophys. Suppl. Ser.* **86**, 119-165.

Landolt, A., 1992. *UBVRI* Photometric Standard Stars in the Magnitude Range of 11.5 < V < 16.0 Around the Celestial Equator. *The Astronomical Journal* **104**, no. 1, 340-371.

Mothé-Diniz, T., Carvano, J. M. and Lazzaro, D., 2003. Distribution of Taxonomic Classes in the Main Belt of Asteroids. *Icarus,* **162***,* 10–21.

Polishook, D. and Brosch, N., 2008. Photometry of Aten Asteroids – More than a Handful of Binaries. *Icarus,* **194***,* 111–124.

Pravec, P., Wolf M. and Šarounová, L., 1998. Lightcurves of 26 Near-Earth Asteroids. *Icarus,* **136***,* 124–153.

Pravec, P. and Harris, A. W., 2000. Fast and slow rotation of asteroids. *Icarus,* **148***,* 12–20.

Pravec, P., Harris, A.W. and Michalowski, T., 2002. Asteroid Rotations. In *Asteroids III* (W.F. Bottke Jr. et al. eds.), pp. 113-122, Univ. of Arizona, Tucson.

Scheeres, D.J., Marzari, F. and Rossi, A., 2004. Evolution of NEO rotation rates due to close encounters with Earth and Venus. *Icarus* **170**, 312-323.

Zappalà, V., Cellino, A., Barucci, A.M., Fulchignoni, M. and Lupishko, D. F., 1990. An analysis of the amplitude-phase relationship among asteroids. *Astron. Astrophys.* **231**, 548-560.

**Tables:**

**Table I:** The observed fields' details: observation date, coordinates (RA & Dec), time span of the nightly observation, the number of images obtained and the number of asteroids in each field.

| Date | RA | Dec | Time span [hours] | Number of Images | Number of asteroids |
|---|---|---|---|---|---|
| Oct 4, 2007 | 02:34:31 | +15:43:51 | 5.91 | 119 | 6 |
| Oct 5, 2007 | 02:32:48 | +15:38:15 | 3.52 | 54 | 7 |
| Oct 6, 2007 | 02:31:11 | +15:33:00 | 5.02 | 86 | 5 |
| Oct 31, 2007 | 01:50:23 | +12:57:30 | 5.62 | 100 | 8 |
| Nov 2, 2007 | 01:47:25 | +12:44:41 | 7.98 | 136 | 11 |
| Nov 3, 2007 | 01:46:00 | +12:38:28 | 4.67 | 81 | 7 |
| Nov 5, 2007 | 02:09:11 | +12:09:02 | 1.39 | 30 | 7 |

**Table II:** Observation circumstances: asteroid name, observation date, nightly time span of the specific observation, the number of images obtained (N), the object's heliocentric (r), and geocentric distances (Δ), the phase angle (α), the Phase Angle Bisector (PAB) ecliptic coordinates ($L_{PAB}$, $B_{PAB}$), and the average magnitude (*mean observed R*) after standard calibration.

| Asteroid name | Date | Time span [hours] | N | r [AU] | Δ [AU] | α [Deg] | $L_{PAB}$ [Deg] | $B_{PAB}$ [Deg] | Mean R [Mag] |
|---|---|---|---|---|---|---|---|---|---|
| *(698) Ernestina* | Nov 6, 2005 | 4.35 | 152 | 2.88 | 1.90 | 3.31 | 36.8 | -0.7 | 14.73 |
| | Nov 7, 2005 | 3.05 | 145 | 2.88 | 1.90 | 3.73 | 36.8 | -0.7 | 14.80 |
| | Jan 21, 2006 | 1.71 | 68 | 2.80 | 2.60 | 20.57 | 43.5 | 2.6 | 15.88 |
| | Jan 22, 2006 | 0.98 | 37 | 2.80 | 2.61 | 20.59 | 43.6 | 2.7 | 16.20 |
| *(8549) Alcide* | Nov 5, 2007 | 1.39 | 30 | 2.65 | 1.66 | 3.40 | 36.1 | -0.5 | 17.80 |
| *(16345) 2391T-3* | Nov 3, 2007 | 4.67 | 81 | 2.69 | 1.71 | 4.50 | 31.4 | 1.1 | 18.57 |
| *(36405) 2000 $OB_{48}$* | Oct 4, 2007 | 5.91 | 107 | 1.97 | 1.05 | 14.90 | 32.0 | 0.2 | 17.50 |
| | Nov 5, 2007 | 1.39 | 24 | 2.01 | 1.02 | 4.25 | 37.3 | -0.6 | 16.87 |
| *(42535) 1995 $VN_9$* | Oct 31, 2007 | 5.62 | 100 | 2.53 | 1.54 | 3.19 | 31.9 | 1.1 | 19.28 |
| | Nov 2, 2007 | 7.98 | 136 | 2.53 | 1.55 | 4.13 | 32.0 | 1.1 | 19.38 |
| | Nov 3, 2007 | 4.67 | 80 | 2.53 | 1.55 | 4.57 | 32.1 | 1.1 | 19.34 |
| *(50891) 2000 $GH_{41}$* | Oct 4, 2007 | 5.91 | 119 | 2.82 | 1.90 | 10.15 | 34.9 | 0.3 | 18.90 |
| *(70517) 1999 $TU_{105}$* | Oct 31, 2007 | 5.62 | 82 | 2.04 | 1.06 | 3.74 | 32.8 | 1.0 | 17.70 |
| *(72963) 2002 $CC_{113}$* | Nov 2, 2007 | 7.98 | 86 | 2.55 | 1.57 | 4.11 | 31.9 | 1.0 | 19.44 |
| | Nov 3, 2007 | 4.67 | 75 | 2.55 | 1.57 | 4.57 | 31.9 | 1.0 | 19.52 |
| *(75598) 2000 $AY_{23}$* | Nov 2, 2007 | 7.98 | 88 | 2.28 | 1.29 | 4.50 | 32.5 | 1.0 | 18.13 |
| *(84939) 2003 $WO_{130}$* | Nov 5, 2007 | 1.39 | 30 | 2.33 | 1.34 | 3.88 | 36.5 | -0.8 | 18.37 |
| *(96404) 1998 $DB_{28}$* | Oct 31, 2007 | 5.62 | 99 | 2.13 | 1.14 | 3.65 | 32.6 | 1.0 | 18.64 |
| *(98176) 2000 $SU_{95}$* | Nov 2, 2007 | 7.98 | 127 | 2.20 | 1.22 | 4.55 | 32.8 | 1.2 | 19.05 |
| *(106836) 2000 $YG_8$* | Oct 31, 2007 | 5.62 | 100 | 2.07 | 1.08 | 3.88 | 32.5 | 0.9 | 18.60 |
| | Nov 2, 2007 | 7.98 | 117 | 2.06 | 1.08 | 5.06 | 32.8 | 0.9 | 18.52 |
| *(106864) 2000 $YL_{27}$* | Oct 5, 2007 | 3.52 | 54 | 2.03 | 1.10 | 13.52 | 31.8 | 0.4 | 18.73 |
| | Oct 6, 2007 | 5.02 | 81 | 2.03 | 1.09 | 13.02 | 31.9 | 0.4 | 18.71 |
| *(129759) 1999 $FU_{60}$* | Nov 5, 2007 | 1.39 | 18 | 2.76 | 1.78 | 3.20 | 36.1 | -0.8 | 20.32 |
| *(129997) 1999 $VH_{28}$* | Nov 2, 2007 | 7.98 | 129 | 2.21 | 1.23 | 4.78 | 32.4 | 1.2 | 18.30 |
| | Nov 3, 2007 | 4.67 | 60 | 2.21 | 1.23 | 5.28 | 32.5 | 1.2 | 18.34 |
| *(132114) 2002 $CE_{224}$* | Oct 4, 2007 | 5.91 | 111 | 2.24 | 1.31 | 12.84 | 32.9 | 0.3 | 19.62 |
| | Oct 5, 2007 | 3.52 | 48 | 2.24 | 1.31 | 12.38 | 33.0 | 0.3 | 19.51 |
| | Oct 6, 2007 | 5.02 | 85 | 2.24 | 1.31 | 11.88 | 33.1 | 0.4 | 19.49 |
| *(166242) 2002 $GL_{12}$* | Oct 5, 2007 | 3.52 | 30 | 2.89 | 1.98 | 9.45 | 34.9 | 0.6 | 19.63 |
| | Oct 6, 2007 | 5.02 | 54 | 2.89 | 1.97 | 9.09 | 35.0 | 0.6 | 20.00 |
| *(166243) 2002 $GL_{13}$* | Nov 5, 2007 | 1.39 | 30 | 2.71 | 1.72 | 3.23 | 36.3 | -0.7 | 19.73 |
| *(168714) 2000 $JQ_{10}$* | Nov 3, 2007 | 4.67 | 72 | 1.83 | 0.84 | 6.54 | 33.3 | 0.9 | 18.47 |
| *(168847) 2000 $UU_{34}$* | Nov 2, 2007 | 7.98 | 127 | 1.99 | 1.01 | 5.43 | 32.7 | 1.1 | 18.77 |
| *(168904) 2000 $WY_{189}$* | Nov 2, 2007 | 7.98 | 111 | 1.83 | 0.84 | 5.62 | 33.5 | 1.1 | 19.00 |
| *(174245) 2002 $RK_{155}$* | Oct 5, 2007 | 3.52 | 26 | 2.74 | 1.82 | 9.91 | 34.3 | 0.4 | 19.99 |
| *2002 $CP_{33}$* | Nov 2, 2007 | 7.98 | 118 | 2.08 | 1.10 | 4.83 | 33.0 | 1.0 | 19.33 |
| | Nov 3, 2007 | 4.67 | 74 | 2.08 | 1.10 | 5.40 | 33.1 | 1.0 | 19.32 |
| *2003 $QU_{50}$* | Oct 4, 2007 | 5.91 | 108 | 2.37 | 1.45 | 12.08 | 33.4 | 0.6 | 19.67 |
| | Oct 5, 2007 | 3.52 | 44 | 2.37 | 1.44 | 11.68 | 33.5 | 0.6 | 19.45 |

**Table II:** Continue.

| Asteroid name | Date | Time span [hours] | N | r [AU] | Δ [AU] | α [Deg] | $L_{PAB}$ [Deg] | $B_{PAB}$ [Deg] | Mean R [Mag] |
|---|---|---|---|---|---|---|---|---|---|
| *2005 EJ_{141}* | Oct 4, 2007 | 5.91 | 107 | 1.89 | 0.96 | 15.32 | 31.0 | 0.5 | 19.99 |
| *2007 TD_{106}* | Oct 6, 2007 | 5.02 | 59 | 2.06 | 1.12 | 12.54 | 31.6 | 0.7 | 20.09 |
| *2007 TF_{5}* | Oct 31, 2007 | 5.62 | 82 | 2.70 | 1.71 | 2.81 | 32.1 | 1.0 | 19.71 |
| *2007 TG_{49}* | Oct 31, 2007 | 5.62 | 100 | 2.68 | 1.70 | 2.97 | 31.9 | 1.4 | 19.33 |
|  | Nov 2, 2007 | 7.98 | 91 | 2.68 | 1.70 | 3.83 | 32.0 | 1.3 | 19.40 |
| *2007 TW_{358}* | Nov 5, 2007 | 1.39 | 30 | 1.81 | 0.82 | 4.93 | 37.5 | -0.5 | 19.39 |
| *2007 TZ_{13}* | Oct 5, 2007 | 3.52 | 45 | 1.87 | 0.93 | 14.57 | 30.5 | 0.2 | 19.02 |
| *2007 UC_{66}* | Oct 31, 2007 | 5.62 | 70 | 2.77 | 1.78 | 2.83 | 31.9 | 1.2 | 20.07 |
| *2007 UM_{110}* | Nov 5, 2007 | 1.39 | 28 | 1.80 | 0.81 | 5.31 | 37.1 | -0.7 | 20.02 |

**Table III:** Analysis results: asteroid's name, period, reliability code, amplitude, absolute magnitude $H_R$, the slope parameter G and range of phase angle. A value of 0.15 was estimated for asteroids without a wide range of phase angle (marked in brackets).

| Name | Period [hours] | Reliability code[+] | Amplitude [mag] | $H_R$ [mag] | G | Phase angle range [deg] |
|---|---|---|---|---|---|---|
| *(698) Ernestina* | 5.0431±0.0002 | 3 | 0.35±0.03 0.26±0.03 | 10.75±0.02 | 0.21±0.02 | 3 - 21 |
| *(16345) 2391T-3* | 2.9±0.2 | 2+ | 0.25±0.05 | 14.89±0.07 | (0.15) | 4.3 |
| *(36405) 2000 OB_{48}* | 6.6±0.3 | 2 | 0.82±0.02 0.62±0.02 | 15.27±0.04 | 0.07±0.04 | 4 – 15 |
| *(42535) 1995 VN_{9}* | 2.37±0.01 | 2+ | 0.3±0.1 | 16.0±0.1 | (0.15) | 3.2 |
| *(75598) 2000 AY_{23}* | 3.05±0.05 | 3 | 0.33±0.03 | 15.41±0.07 | (0.15) | 4.5 |
| *(106836) 2000 YG_{8}* | 20.1±0.2 | 2 | 0.65±0.03 | 16.53±0.08 | (0.15) | 4.5 |
| *(129997) 1999 VH_{28}* | 8.8±0.2 | 2 | 0.25±0.03 | 15.74±0.07 | (0.15) | 5.0 |
| *(132114) 2002 CE_{224}* | 5.672±0.003 | 3 | 1.25±0.05 | 16.7±0.2 | (0.15) | 12.4 |
| *(168714) 2000 JQ_{10}* | 2.5±0.2 | 2+ | 0.15±0.05 | 17.01±0.09 | (0.15) | 6.6 |
| *(168847) 2000 UU_{34}* | 2.9±0.1 | 2+ | 0.30±0.05 | 16.79±0.08 | (0.15) | 5.6 |
| *(168904) 2000 WY_{189}* | 3.335±0.009 | 3 | 0.36±0.05 | 17.58±0.09 | (0.15) | 6.2 |
| *2002 CP_{33}* | 4.78±0.02 | 3 | 0.8±0.1 | 17.2±0.1 | (0.15) | 5.1 |
| *2003 QU_{50}* | 4.14±0.01 | 3 | 0.6±0.1 | 16.3±0.2 | (0.15) | 12.1 |
| *2007 TF_{5}* | 2.5±0.2 | 2+ | 0.4±0.1 | 16.1±0.1 | (0.15) | 2.8 |

[+] A reliability code of 2+ stands for results based on full coverage of the period but with low amplitude compared to the S/N.

**Table IV:** Photometry results for asteroids with inconclusive analysis: asteroid name, estimation of minimal periodicity, minimal variability, and absolute magnitude $H_R$.

| Name | Minimal period [hours] | Minimal variability [mag] | Absolute mag. $H_R$ [mag] |
|---|---|---|---|
| (8549) Alcide | 2.78 | 0.25 | 14.3±0.4 |
| (70517) 1999 $TU_{105}$ | 11.24 | 0.21 | 15.7±0.4 |
| (106864) 2000 $YL_{27}$ | 4.5 | 0.30 | 16.2±0.4 |
| (166243) 2002 $GL_{13}$ | 2.78 | 0.2 | 16.1±0.5 |
| 2005 $EJ_{141}$ | 21.0 | 0.5 | 17.9±0.5 |

**Table V:** Photometry results for asteroids without period analysis: asteroid name, number of data points (N), absolute magnitude ($H_R$) assuming G=0.15.

| Asteroid Name | Data points | Absolute magnitude $H_R$ [mag] |
|---|---|---|
| (50891) 2000 $GH_{41}$ | 119 | 14.6±0.5 |
| (72963) 2002 $CC_{113}$ | 161 | 16.1±0.5 |
| (84939) 2003 $WO_{130}$ | 30 | 15.5±0.5 |
| (96404) 1998 $DB_{28}$ | 99 | 16.4±0.5 |
| (98176) 2000 $SU_{95}$ | 127 | 16.5±0.5 |
| (129759) 1999 $FU_{60}$ | 18 | 16.6±0.5 |
| (166242) 2002 $GL_{12}$ | 84 | 15.6±0.5 |
| (174245) 2002 $RK_{155}$ | 26 | 16.2±0.5 |
| 2007 $TD_{106}$ | 59 | 17.7±0.5 |
| 2007 $TG_{49}$ | 191 | 15.8±0.5 |
| 2007 $TW_{358}$ | 30 | 18.1±0.5 |
| 2007 $TZ_{13}$ | 45 | 17.0±0.5 |
| 2007 $UC_{66}$ | 70 | 16.4±0.5 |
| 2007 $UM_{110}$ | 28 | 18.8±0.5 |

**Table VI:**: assumed albedos and calculated diameters for each asteroid based on their semi major axis.

Diameters' errors are based only on the errors of $H_R$ and are not based on the albedo estimate.

| Name | Semi major axis [AU] | Group/Family | Assumed albedo | Diameter [km] |
|---|---|---|---|---|
| *(698) Ernestina* | 2.870 | Outer MB | 0.058 | 31.8±0.3 |
| *(16345) 2391T-3* | 2.918 | Outer MB | 0.058 | 4.7±0.1 |
| *(36405) 2000 OB$_{48}$* | 2.285 | Flora | 0.24 | 1.95±0.04 |
| *(42535) 1995 VN$_9$* | 2.515 | Inner MB | 0.18 | 1.61±0.07 |
| *(75598) 2000 AY$_{23}$* | 2.633 | Central MB | 0.1 | 2.83±0.09 |
| *(106836) 2000 YG$_8$* | 2.381 | Inner MB | 0.18 | 1.26±0.05 |
| *(129997) 1999 VH$_{28}$* | 2.548 | Inner MB | 0.18 | 1.81±0.06 |
| *(132114) 2002 CE$_{224}$* | 2.352 | Inner MB | 0.18 | 1.2±0.1 |
| *(168714) 2000 JQ$_{10}$* | 2.210 | Inner MB | 0.18 | 1.01±0.04 |
| *(168847) 2000 UU$_{34}$* | 2.311 | Inner MB | 0.18 | 1.12±0.04 |
| *(168904) 2000 WY$_{189}$* | 2.321 | Flora | 0.24 | 0.67±0.03 |
| *2002 CP$_{33}$* | 2.200 | Flora | 0.24 | 0.80±0.04 |
| *2003 QU$_{50}$* | 2.390 | Vestoid | 0.18 | 1.4±0.1 |
| *2007 TF$_5$* | 2.607 | Central MB | 0.1 | 2.1±0.1 |

**Figures:**

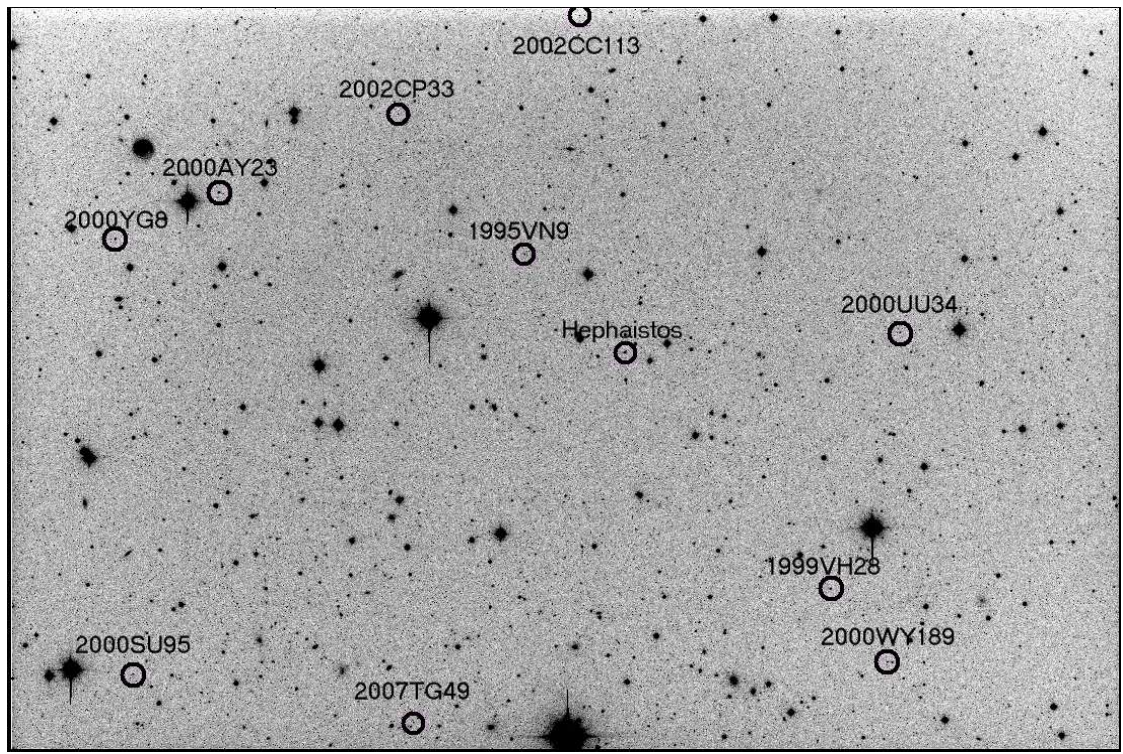

Fig. 1: A *C18* image obtained on November 2, 2007 (JD=2454407.412). A record of 11 asteroids appear in this field. Each asteroid is circled and labeled by the object name.

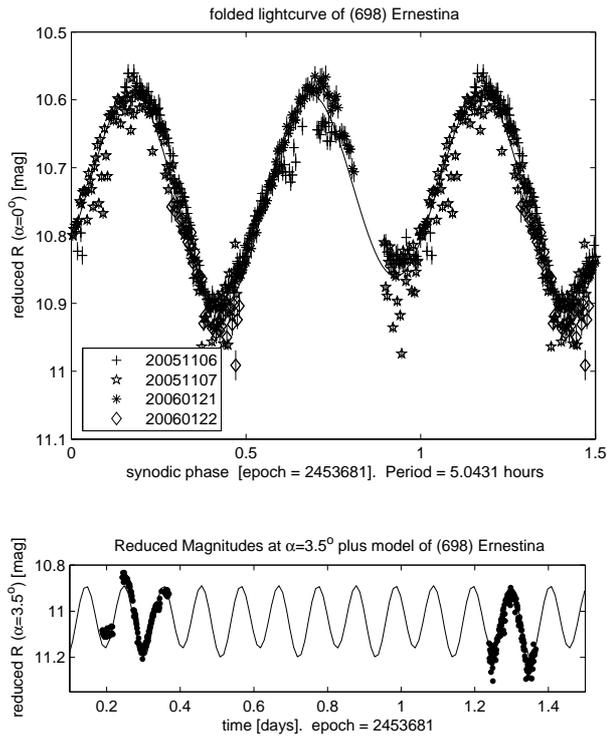

Fig. 2: *(698) Ernestina* lightcurve, folded with a period of 5.0431 hours (top panel). The reduced data points of November 2005 with the model are exhibited in the lower panel. Here and in subsequent folded lightcurve plots (top panels) the individual points carry error bars; these are sometimes too small to see. Lower panels were plotted without error bars for display reasons.

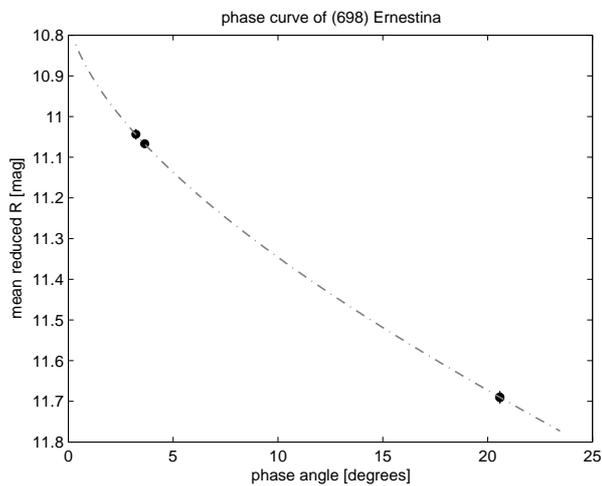

Fig. 3: Phase curve of *(698) Ernestina*. The grey dashed line represents the *H-G* model.

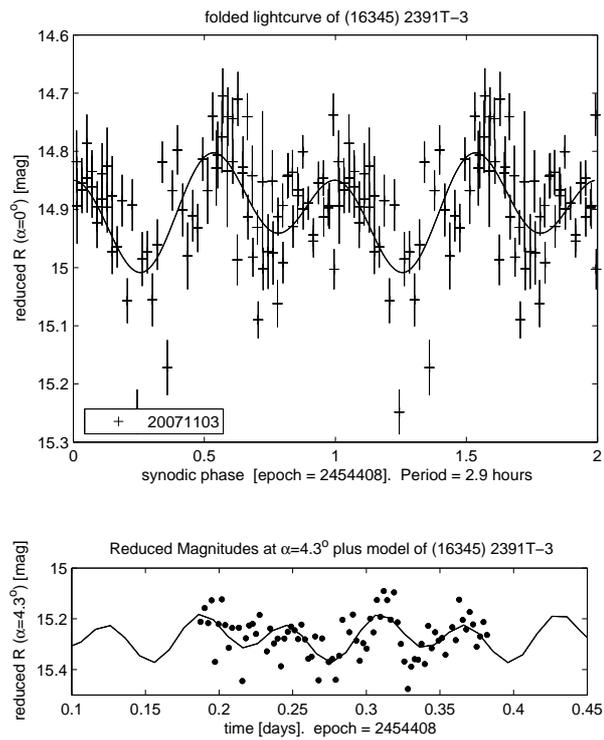

Fig. 4: *(16345) 2391T-3* lightcurve, folded with a period of 2.9 hours (top panel). The reduced data points, with the model superposed as a solid line, are exhibited in the lower panel.

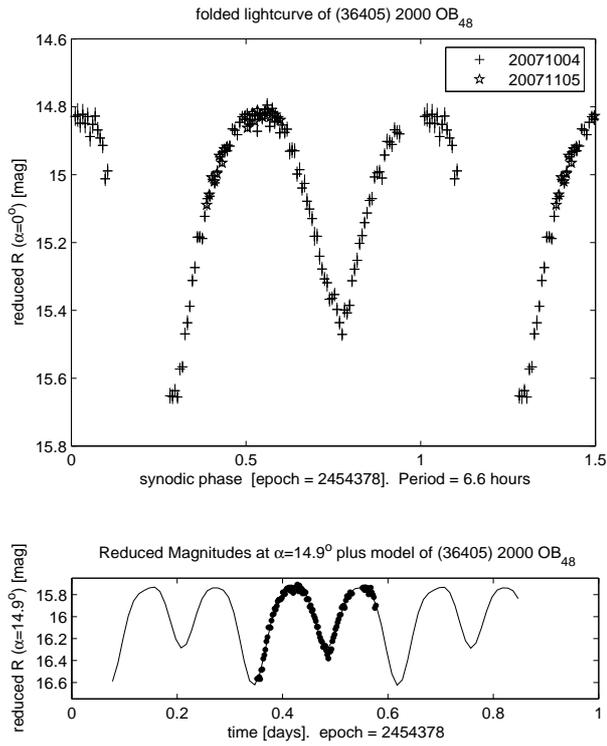

Fig. 5: *(36405) 2000 OB₄₈* lightcurve, folded with a period of 6.6 hours (top panel). The reduced data points of October 4 with the model are exhibited in the lower panel.

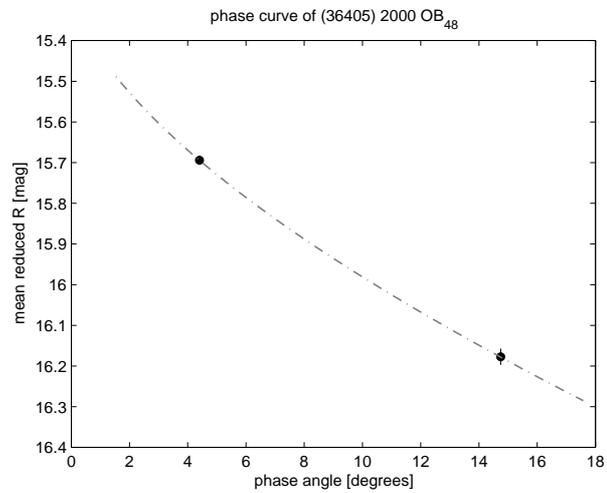

Fig. 6: Phase curve of *(36405) 2000 OB₄₈*. The grey dashed line represents the *H-G* model.

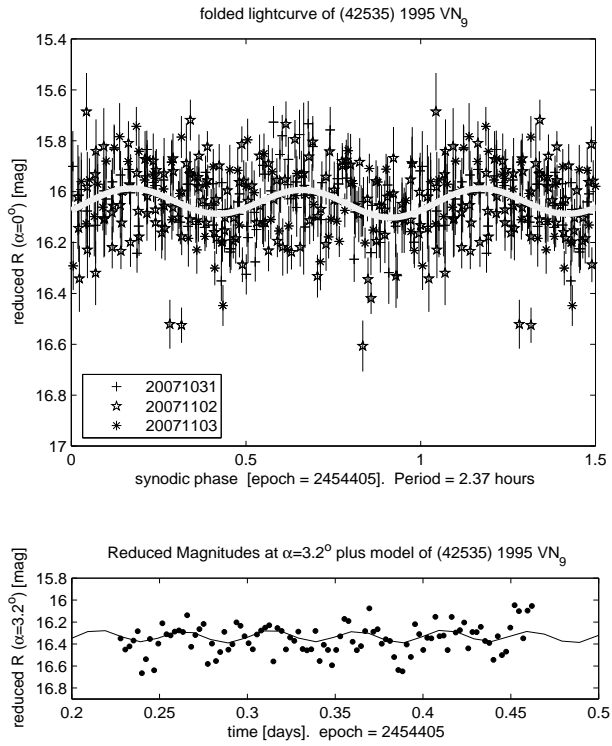

Fig. 7: *(42535) 1995 VN₉* lightcurve, folded with a period of 2.37 hours (top panel) with the model (grey line). The reduced data points for the night of October 31, together with the fitted model are shown in the lower panel.

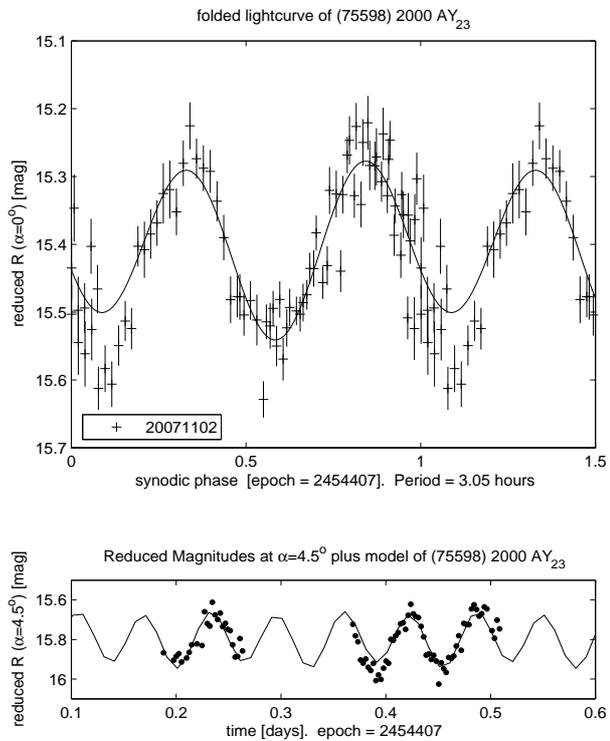

Fig. 8: *(75598) 2000 AY₂₃* lightcurve, folded with a period of 3.05 hours (top panel). The reduced data points are exhibited together with the fitted model in the lower panel.

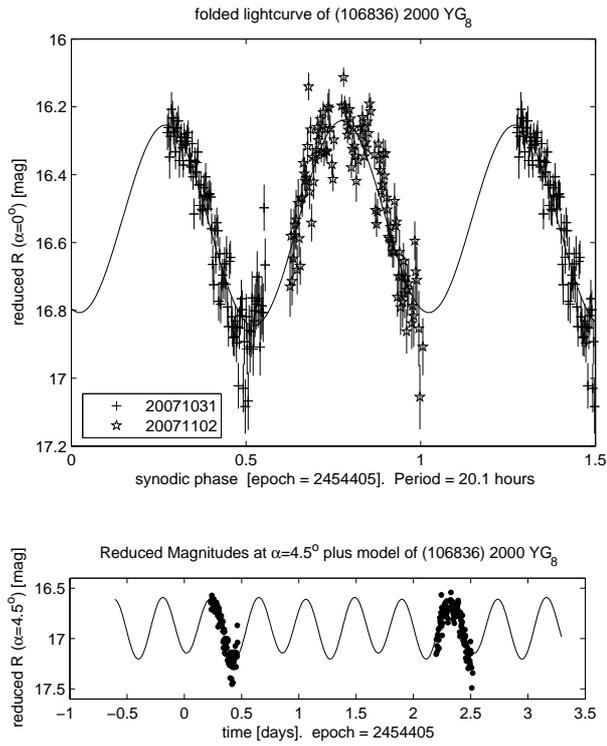

Fig. 9: *(106836) 2000 YG$_8$* lightcurve, folded with a period of 20.1 hours (top panel). The reduced data points are exhibited together with the fitted model in the lower panel.

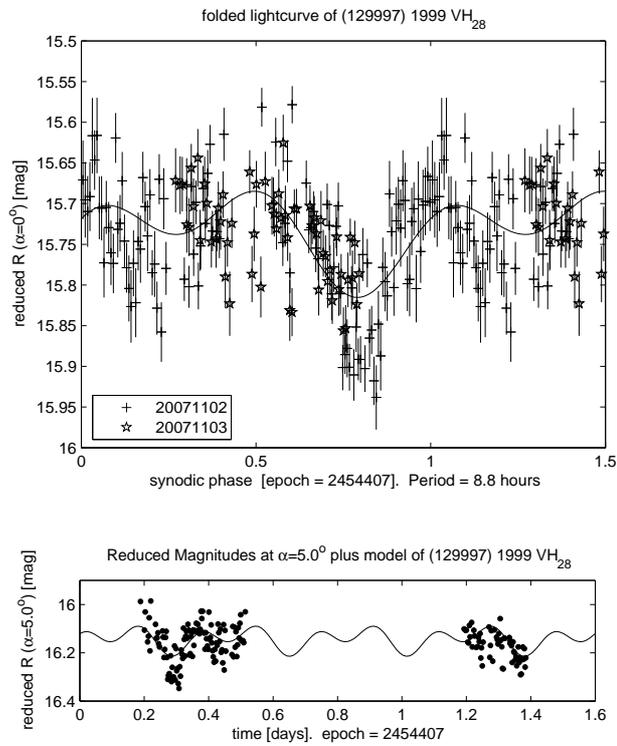

Fig. 10: *(129997) 1999 VH$_{28}$* lightcurve, folded with a period of 8.8 hours (top panel). The reduced data points with the model are exhibited in the lower panel.

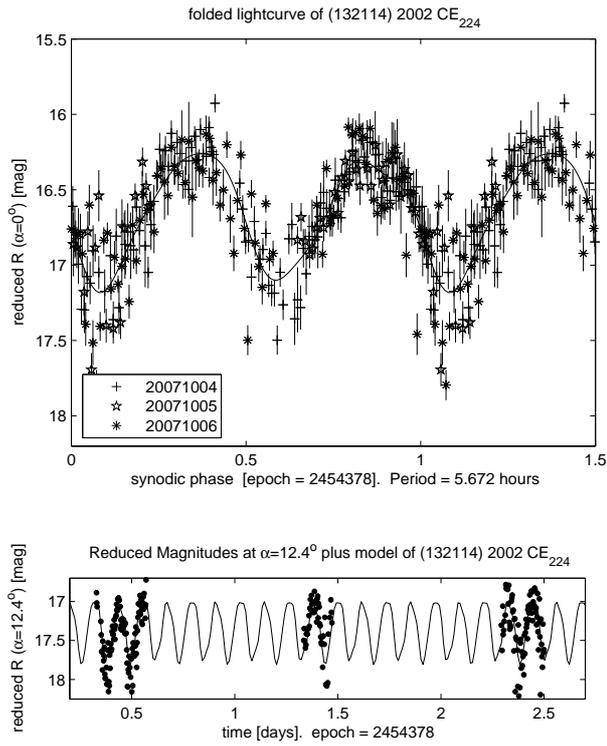

Fig. 11: *(132114) 2002 CE$_{224}$* lightcurve, folded with a period of 5.672 hours (top panel). The reduced data points with the model are exhibited in the lower panel.

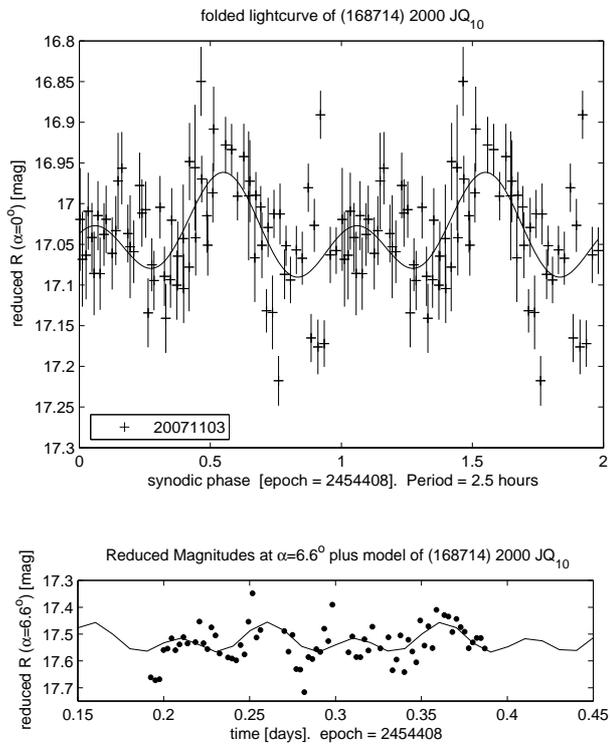

Fig. 12: *(168714) 2000 JQ$_{10}$* lightcurve, folded with a period of 2.5 hours (top panel). The reduced data points with the model are exhibited in the lower panel.

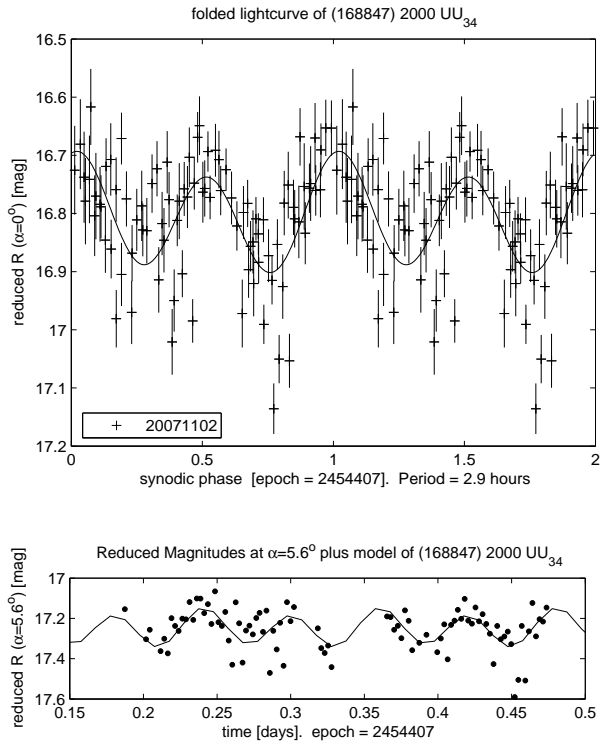

Fig. 13: *(168847) 2000 UU$_{34}$* lightcurve, folded with a period of 2.9 hours (top panel). The reduced data points with the model are exhibited in the lower panel.

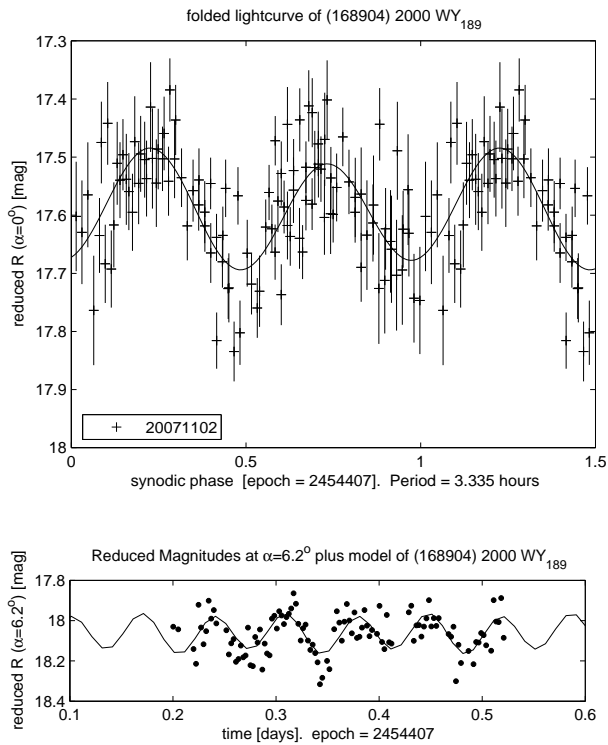

Fig. 14: *(168904) 2000 WY$_{189}$* lightcurve, folded with a period of 3.335 hours (top panel). The reduced data points with the model are exhibited in the lower panel.

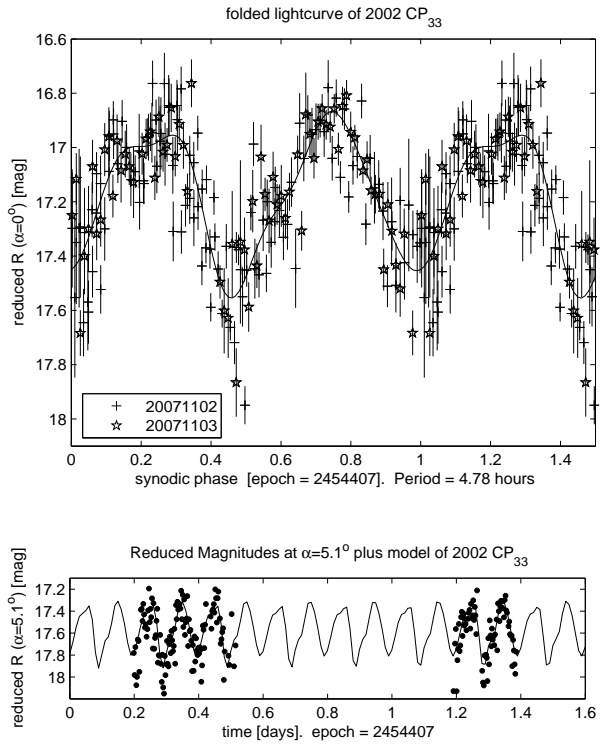

Fig. 15: *2002 CP$_{33}$* lightcurve, folded with a period of 4.78 hours (top panel). The reduced data points with the model are exhibited in the lower panel.

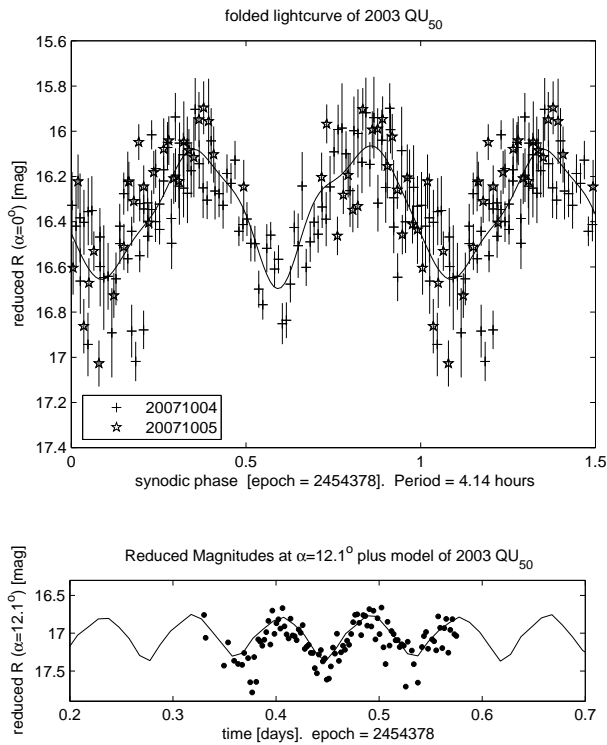

Fig. 16: *2003 QU$_{50}$* lightcurve, folded with a period of 4.14 hours (top panel). The reduced data points of October 4 with the model are exhibited in the lower panel.

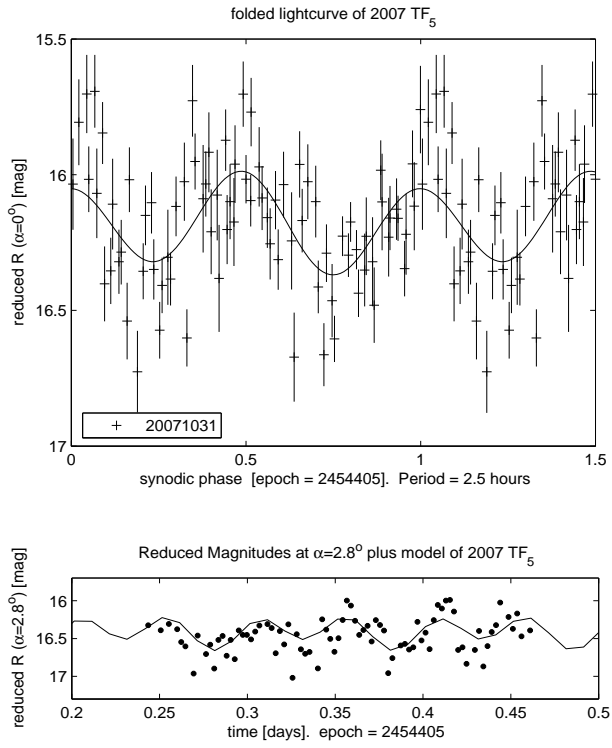

Fig. 17: *2007 TF$_5$* lightcurve, folded with a period of 2.5 hours (top panel). The reduced data points with the model are exhibited in the lower panel.

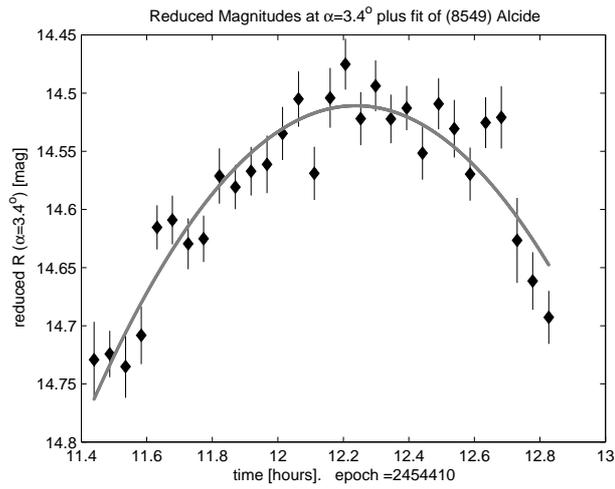

Fig. 18: *(8549) Alcide* lightcurve after reduction. A 2$^{nd}$ order polynomial fit is marked by the grey line.

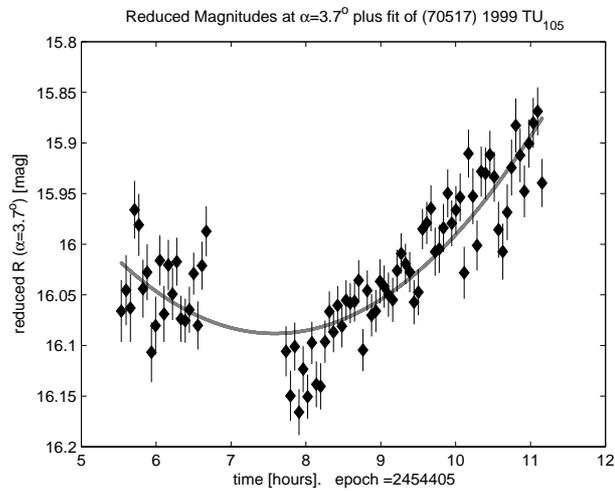

Fig. 19: *(70517) 1999 TU$_{105}$* lightcurve after reduction. A 2$^{nd}$ order polynomial fit is marked by the grey line.

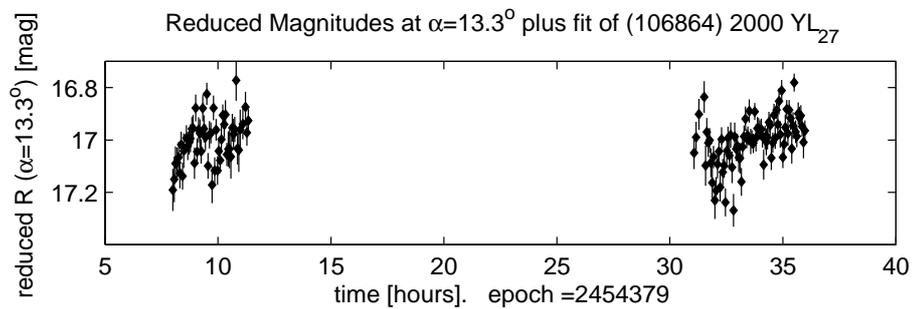

Fig. 20: *(106864) 2000 YL$_{27}$* lightcurve after reduction on October 5 & 6.

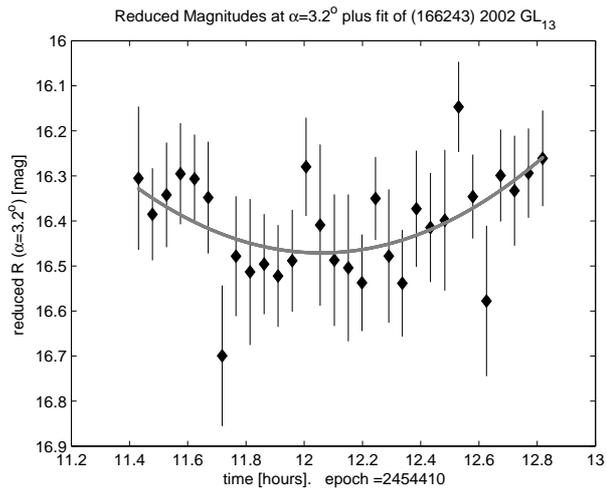

Fig. 21: *(166243) 2002 GL$_{13}$* lightcurve after reduction. A 2$^{nd}$ order polynomial fit is marked by the grey line.

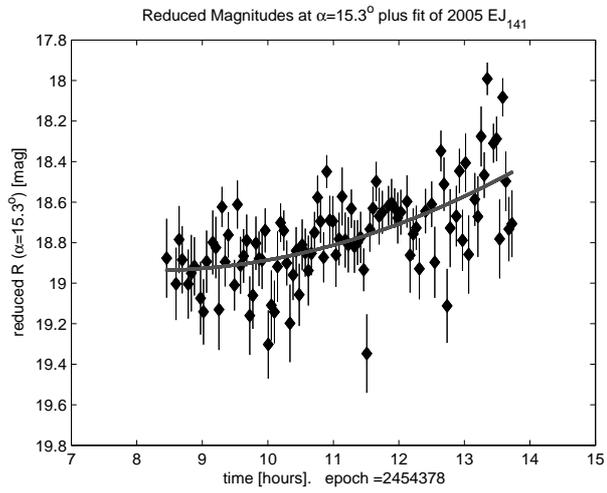

Fig. 22: *2005 EJ$_{141}$* lightcurve after reduction. A 2$^{nd}$ order polynomial fit is marked by the grey line.

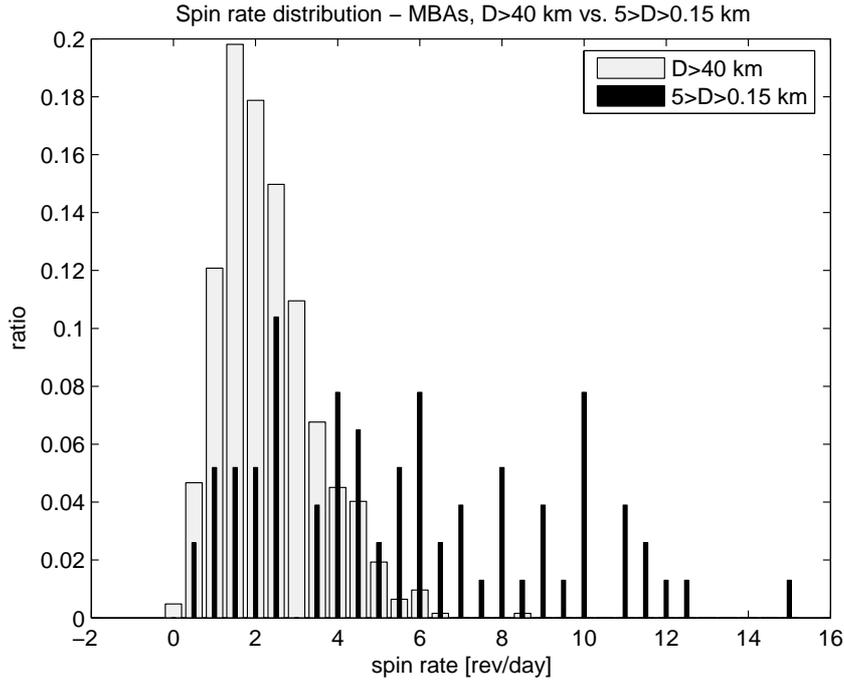

Fig. 23: Comparison between spin rate distributions of MBAs with diameter of D>40 km (grey) and MBAs with diameter of 5>D>0.15 km (black). The distribution is based on 621 periods from the L group of MBAs and 77 periods from the VS group of MBAs.

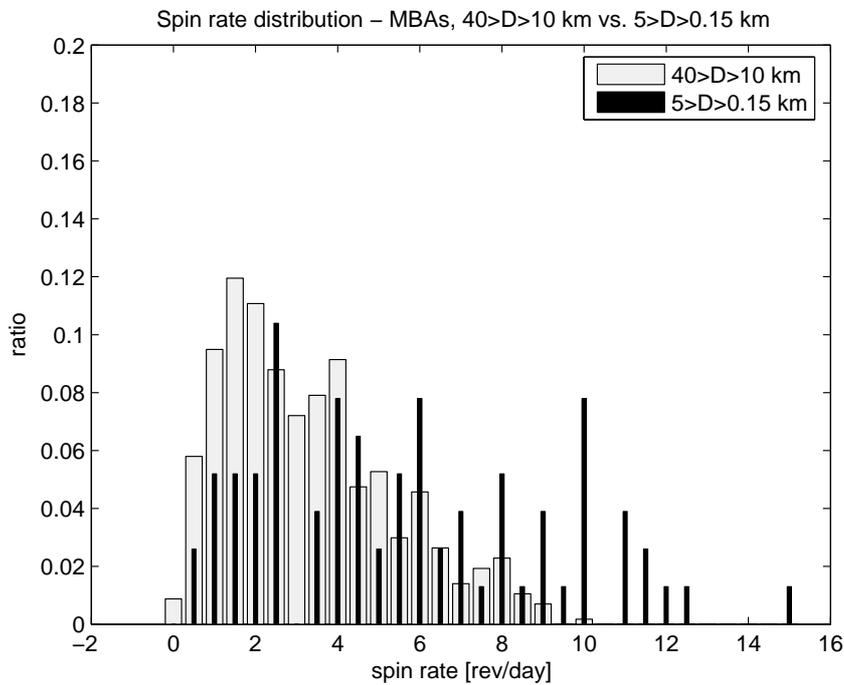

Fig. 24: Comparison between spin rate distributions of MBAs with diameter of 40>D>10 km (grey) and MBAs with diameter of 5>D>0.15 km (black). The distribution is based on 569 periods from the M group of MBAs and 77 periods from the VS group of MBAs.

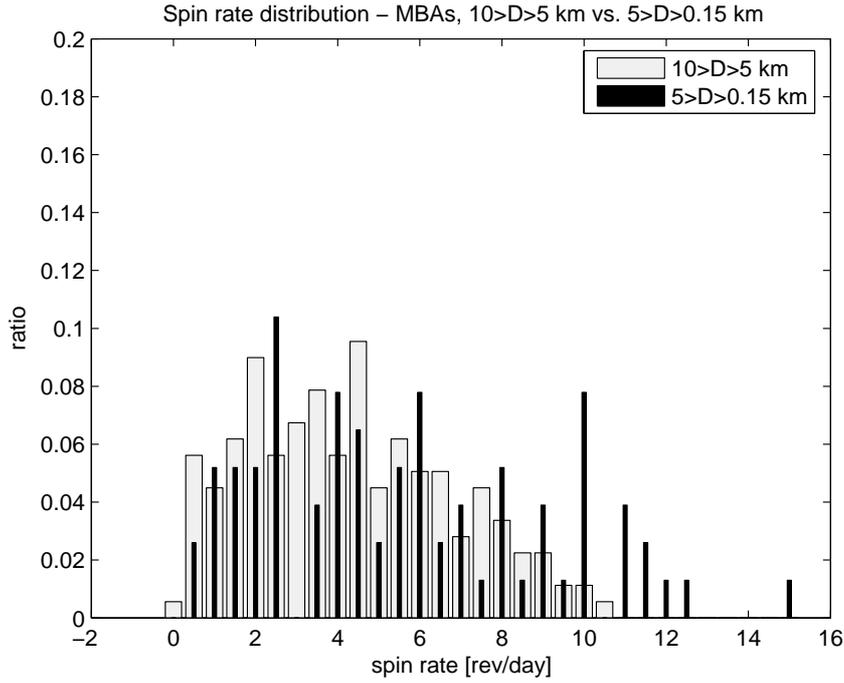

Fig. 25: Comparison between spin rate distributions of MBAs with diameter 10>D>5 km (grey) and MBAs with diameter of 5>D>0.15 km (black). The distribution is based on 178 periods from the S group of MBAs and 77 periods from the VS group of MBAs.

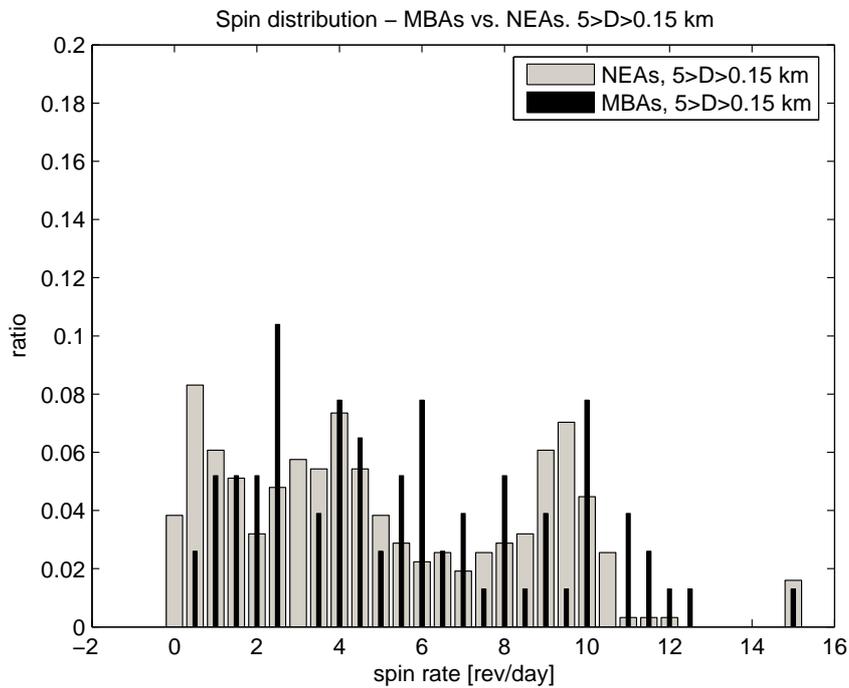

Fig. 26: Comparison between spin rate distributions of NEAs (grey) and MBAs (black) both with diameter of 5>D>0.15 km. The distribution is based on 286 periods of NEAs and 77 periods from the VS group of MBAs.

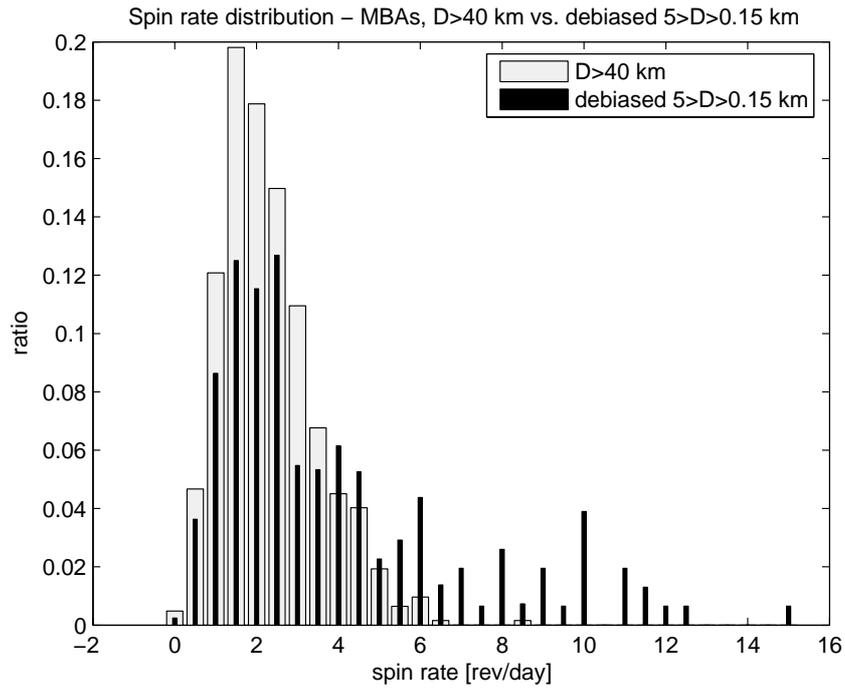

Fig. 27: Comparison between spin rate distributions of MBAs with diameter of D>40 km (grey) and debiased spin rate distribution of MBAs with diameter of 5>D>0.15 km (black). The debias is based on the distribution of the large MBAs. See text for further details.